\documentclass[submission,copyright,creativecommons]{eptcs}
\providecommand{\event}{FMAS 2026} % Name of the event you are submitting to

\usepackage{iftex}

\ifpdf
  \usepackage[strings]{underscore}         % Only needed if you use pdflatex.
  \usepackage[T1]{fontenc}        % Recommended with pdflatex
\else
  \usepackage{breakurl}           % Not needed if you use pdflatex only.
\fi

\usepackage{graphicx}

\usepackage{amsmath}
\usepackage{amssymb}

\def\markDot{\bullet}
\newcommand{\preset}[1]{^\markDot{}#1}

\def\toolformat{\textsc}
\def\workcraft{\toolformat{Workcraft}}
\def\petrify{\toolformat{Petrify}}
\def\mpsat{\toolformat{MPSat}}
\def\reach{\toolformat{REACH}}

\newcommand{\transitionName}[1]{\texttt{\text{#1}}}

\usepackage{enumitem}
\setlist[itemize]{noitemsep}

\usepackage{algorithm} 
\usepackage{algpseudocode} 

\usepackage{booktabs}
\usepackage{multirow}

\usepackage{bm}

\title{From Constraints to Causality: Petri net Encodings for Explaining Binary Neural Network Inference}
\def\titlerunning{From Constraints to Causality: Petri net Encodings for Explaining BNN Inference}
\author{ 
    Alex Chan$^*$, 
    Mohamed Tarraf$^*$,
    Alex Yakovlev 
    and Rishad Shafik
    \\ \small{$^*$ indicates equal contribution}
    \institute{
        School of Engineering, Newcastle University, Newcastle upon Tyne, United Kingdom
    }
    \vspace{-6px}
}
\def\authorrunning{A. Chan, M. Tarraf, A. Yakovlev and R. Shafik}

\begin{document}

\maketitle

%%%%%%%%%%%%%%%%%%%%%%%%%%%%%%%%%%%%%%%%%%%%%%%%%%%%%%%%%%%%%%%%%%%%%
%%%%%%%%%%%%%%%%%%%%%%%%%%%%%%%%%%%%%%%%%%%%%%%%%%%%%%%%%%%%%%%%%%%%%
%%%%%%%%%%%%%%%%%%%%%%%%%%%%%%%%%%%%%%%%%%%%%%%%%%%%%%%%%%%%%%%%%%%%%

\begin{abstract}
Binary Neural Networks (BNNs) are attractive for resource-constrained hardware and on-device components in autonomous systems. However, their discrete, non-linear behaviour makes them difficult to analyse and trust in safety-relevant settings. Existing approaches based on SAT and Satisfiability Modulo Theories can verify a BNN’s input-output properties, but treat it as a monolithic constraint system, which reveal only whether a property holds and not how the computation reaches its result. In this paper, we present an early work that provides a complementary, discrete-event view. We give a compositional translation from the inference semantics of a fixed, trained BNN to a 1-safe Petri net (PN), where each computation step is an explicit transition, so that a firing sequence is a causal trace of the BNN’s decision. On a worked example, we establish faithfulness via exhaustive input-output equivalence against a reference implementation, and show how contrasting firing sequences localises where a single-feature change alters the prediction. Finally, we outline two directions this opens: a correctness-of-translation theorem, and a causal-trace analysis of where two trained BNNs diverge in their decisions, a structural question that constraint-based verifiers cannot express.
\end{abstract}

%%%%%%%%%%%%%%%%%%%%%%%%%%%%%%%%%%%%%%%%%%%%%%%%%%%%%%%%%%%%%%%%%%%%%
%%%%%%%%%%%%%%%%%%%%%%%%%%%%%%%%%%%%%%%%%%%%%%%%%%%%%%%%%%%%%%%%%%%%%
%%%%%%%%%%%%%%%%%%%%%%%%%%%%%%%%%%%%%%%%%%%%%%%%%%%%%%%%%%%%%%%%%%%%%

\section{Introduction}\label{section:introduction}

Binary Neural Networks (BNNs)~\cite{courbariaux2016bnn} constrain weights and activations to binary values, effectively reducing inference to bitwise operations and thresholding. 
This has driven the BNN's adoption onto resource-constrained hardware~\cite{Umuroglu_2017} and in safety-relevant applications, such as wearable activity recognition~\cite{hosseini2019minimizing},
time-series health monitoring~\cite{2023-ecai-pu_et_al-arrhythmia_classification_via_bnn}, 
and on-device components within autonomous systems~\cite{frickenstein2020binary},
which demand a tight balance between accuracy and memory footprint. 

However, integrating artificial intelligence components into such operational systems raises dependability concerns for engineers, who are responsible for ensuring that the system is safe, reliable and assured.
For example, consider a BNN that is deployed as an on-device perception component in an autonomous driving system, such as a drivable-area classifier~\cite{frickenstein2020binary}: When an unexpected classification is produced, an engineer must determine the cause by analysing which inputs and internal computations led to that decision, and localise the fault to evaluate whether it is a systematic flaw in the model or an isolated case.
In practice, this often requires mechanisms that not only verify properties of machine learning (ML) models, but also trace their internal computational steps during execution.

Despite this, BNNs largely remain opaque from the standpoint of verification, as their behaviour is not directly readable from the parameter matrices~\cite{leblanc_Seeking_Interpretability_Explainability_Binary_Activated_Neural_Networks}. Existing methods either treat the network as a feature-attribution black box~\cite{2017-neurips-lundberg_scott_lee-SHAP,2016-kdd-ribeiro_singh_guestrin-LIME} or as a monolithic constraint system for SAT-based and Satisfiability Modulo Theories (SMT)-based robustness checks~\cite{2018-aaai-nina_et_al-bnn_verification,2020-neurips-jia-exact_verification_of_bnns}. 
While these methods are powerful, they only analyse the network as a constraint system, and provide limited visibility into the operational structure of the computation itself. 
As a result, they may offer guarantees of outcomes, but at the cost of not revealing how internal events and dependencies actually contribute to those outcomes.

We address this gap by giving the BNN a discrete-event counterpart using a Petri net (PN)~\cite{1989-murata-ieee_journal-petrinets}, which is a well-established formalism for representing concurrent and event-driven systems.
Unlike computational-graph formats, such as ONNX~\cite{2019-onnx-github-repo}, which describe a network’s operations but leave their execution to an external runtime, a PN is itself an executable semantics that can be analysed for reachability.
Other formalisms like standard compositions of automata typically represent concurrent steps by interleaving them, while PNs represent concurrency as independent transitions.
By representing thresholding steps and activations as event-based transitions that are connected through token flows, the resulting model exposes the BNN's causal dependencies, ordering constraints, and concurrent patterns that can be extracted from the PN's unfolding semantics~\cite{2004-khomenko-acsd-mpsat}.

This representation should not be seen as a replacement for existing verification techniques, but rather a complementary methodology. Input–output properties checked by SAT-based and SMT-based verifiers~\cite{2018-aaai-nina_et_al-bnn_verification,2020-neurips-jia-exact_verification_of_bnns} can also be expressed as 
% reachability queries over the PN (Section~\ref{section:validation:equivalence}), 
reachability queries over the PN as shown in Section~\ref{section:validation:equivalence},
though dedicated BNN verifiers remain better suited to check such properties at scale. 
Instead, the PN provides what these verifiers do not cover: a firing sequence exposing the causal structure of the BNN’s decision, rather than just an assignment of input values.
From the perspective of model designers and verification engineers, who have access to the network’s architecture, such causal models can support debugging unexpected model behaviour, tracing how faults propagate through inference pipelines, and reasoning about ML components within larger dependable system architectures.
Thus, our contributions include:

\begin{itemize}
    \item A compositional translation from the inference semantics of a fixed, trained BNN to a 1-safe PN, in which each computation step is a transition (Section~\ref{section:methodology}).
    \item An exhaustive faithfulness evaluation on a worked example that combines input--output equivalence with a reachability-based determinacy check (Section~\ref{section:validation:equivalence}).
    \item A demonstration of how contrasting firing sequences localises a decision change to the responsible neuron, which expose the causal structure that constraint-based verifiers cannot (Section~\ref{section:validation:trace}).
\end{itemize}

\vspace{-5px}
%%%%%%%%%%%%%%%%%%%%%%%%%%%%%%%%%%%%%%%%%%%%%%%%%%%%%%%%%%%%%%%%%%%%%
%%%%%%%%%%%%%%%%%%%%%%%%%%%%%%%%%%%%%%%%%%%%%%%%%%%%%%%%%%%%%%%%%%%%%
%%%%%%%%%%%%%%%%%%%%%%%%%%%%%%%%%%%%%%%%%%%%%%%%%%%%%%%%%%%%%%%%%%%%%
\section{Background}\label{section:background}

\subsection{Binary Neural Networks}\label{section:background:BNN}

BNNs~\cite{courbariaux2016bnn} are deep neural networks~\cite{lecun2015deep} whose weights and activations are binary, i.e. in $\{-1,+1\}$. 
Formally, a BNN is a tuple 
$N = (L, k, W_b, \sigma)$ where $L$ is the number of layers, indexed $l = 1, \ldots, L$; 
$k = (k_1, \ldots, k_L)$ is the number of neurons per layer,  with $k_1$ and $k_L$ being the input and output dimensions; 
$W_b = \{W_b^{(l)}\}_{l=2}^{L}$ is the set of binary weight matrices $W_b^{(l)} \in \{-1,+1\}^{k_l \times k_{l-1}}$; 
and 
$\sigma$ is the Sign activation function such that $\mathrm{Sign}(x) = +1$ if $x \geq 0$ and $-1$ otherwise.

BNNs follow the typical standard of a feed-forward architecture containing input, hidden, and output layers. At inference, each neuron computes a weighted sum using binary weights, and applies the \textit{Sign} function to produce a binary activation. The output layer then aggregates the (active) neuron outputs to generate the final prediction. 

\subsection{Petri nets}\label{section:background:PN}

PNs \cite{1989-murata-ieee_journal-petrinets} are a simple, yet powerful, mathematical model that can be used to specify and analyze many concurrent systems including distributed systems, asynchronous circuits~\cite{1985-yakovlev-petrinets_workshop-stg} and logic-based ML models~\cite{2021-async-wheeldon_yakovlev_shafik-self_timed_reinforcement_learning_using_tm, 2024-chan_wheeldon_shafik_yakovlev-event_driven_tsetlin_machines_using_petri_nets}.
Formally, PNs can be defined as a tuple $G = (P, T, F, M)$ where
$P$ is the finite set of places (drawn as circles);
$T$ is the finite set of transitions (drawn as squares) such that $P \cap T = \emptyset$;
$F \subseteq (P \times T) \cup (T \times P)$ is the set of arcs (drawn as arrows) from places to transitions and from transitions to places; and
$M: P \rightarrow \{0, 1, 2, \ldots\}$ is a multi-set of places, i.e. a marking (drawn as black dots or numbers inside places), that represents the PN's state with $M_0$ being the initial marking.

For this paper, we consider arcs of weight $1$, where $W: (P \times T) \cup (T \times P) \rightarrow \{0,1\}$ is the arc-weight function such that $W(x,y) = 1$ if $(x,y) \in F$ and $W(x,y) = 0$ otherwise. 
A transition $t \in T$ is said to be enabled at marking $M$ if every input place holds a token, i.e. $\forall p \in \preset{t}:\; M(p) \geq 1$, where $\preset{t} = \{p \in P \mid (p,t) \in F\}$ is the preset of $t$. 
When $t$ is fired, the marking changes from $M$ to $M'$, i.e. $M'(p) = M(p) - W(p,t) + W(t,p)$ for all $p \in P$, and a sequence of such firings forms a firing sequence.

PNs are also supported by tools like \workcraft{}~\cite{2006-workcraft-website}, which has a graphical front-end to design, edit and simulate graph-based models, with a well-established backend including \petrify{}~\cite{1997-cortadella-ieice-petrify} and \mpsat{}~\cite{2004-khomenko-acsd-mpsat} for formal verification.
In this particular work, we used \workcraft{} to design the BNN's components to leverage \mpsat{}'s \reach{} expressions to verify custom properties at the level of BNNs. 

%%%%%%%%%%%%%%%%%%%%%%%%%%%%%%%%%%%%%%%%%%%%%%%%%%%%%%%%%%%%%%%%%%%%%
%%%%%%%%%%%%%%%%%%%%%%%%%%%%%%%%%%%%%%%%%%%%%%%%%%%%%%%%%%%%%%%%%%%%%
%%%%%%%%%%%%%%%%%%%%%%%%%%%%%%%%%%%%%%%%%%%%%%%%%%%%%%%%%%%%%%%%%%%%%

\section{Petri net Encoding of Binary Neural Networks}\label{section:methodology}

In this section, we present our PN encoding of BNNs with fully connected (dense) layers, which produces a separate PN model of a given trained BNN. The BNN itself is unchanged and the PN mirrors the BNN's structure, with one segment per operation, so that each step of the BNN’s inference corresponds to identifiable transitions in the PN.

\subsection{Input Loading}\label{section:methodology:input-loading}

Since each value is represented by its own place, our encoding assumes binary-valued inputs, i.e. $x \in \{-1, +1\}^D$, where $D = k_1$ is the number of input features. Features in other formats are mapped into this domain during preprocessing, e.g. for Boolean data, a logical $1$ maps to $+1$ and a logical $0$ to $-1$.

\begin{figure}[ht]
    \centering
    \vspace{-5px}
    \includegraphics[width=0.5\linewidth]{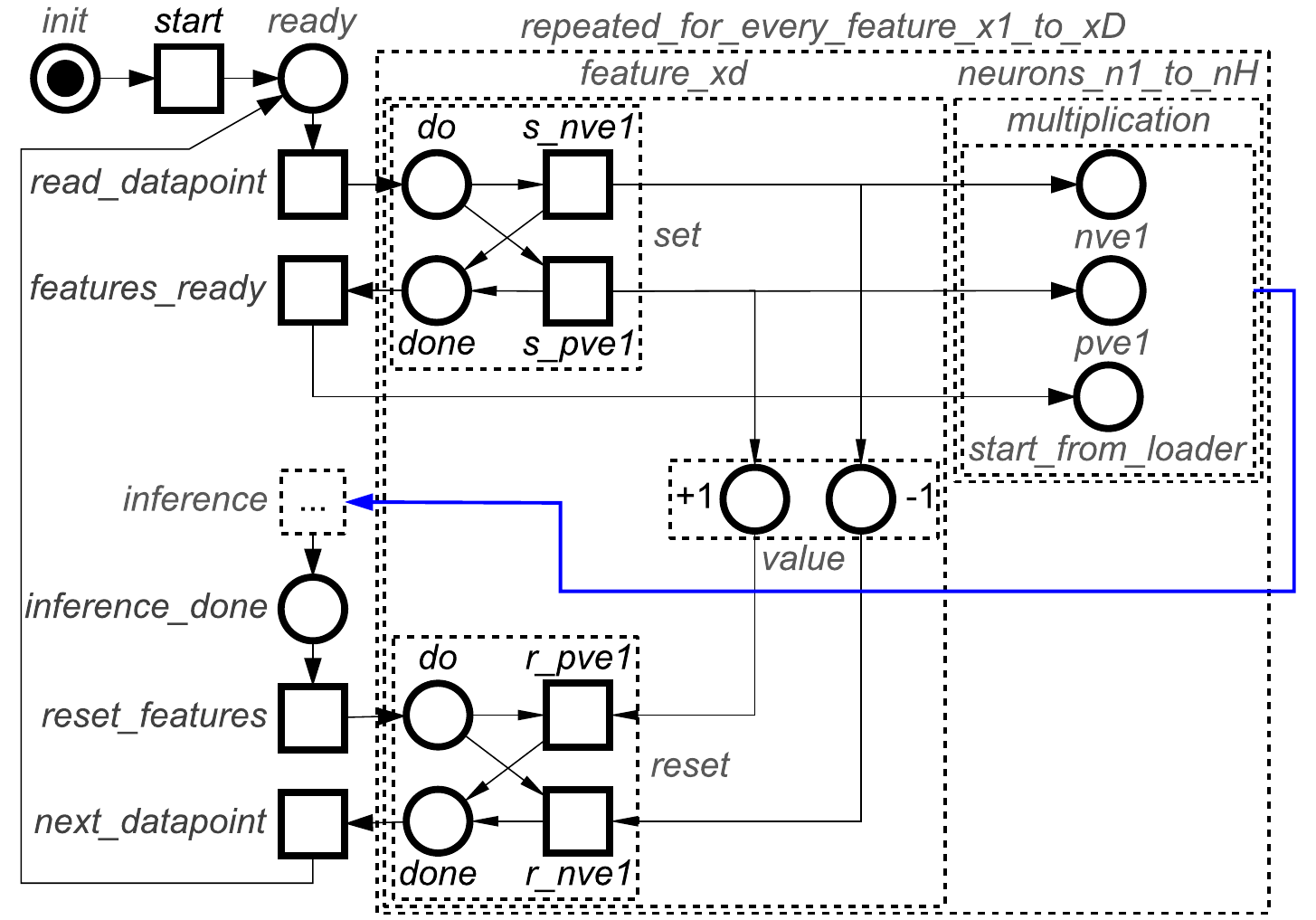}
    \vspace{-2.5px}
    \caption{
        PN segment for loading the datapoints. The \textit{inference} block abstracts the pipeline of Figure~\ref{fig:inference} with the blue arrow indicating the multiplication places it consumes, while the other dashed boxes group the places and transitions repeated for each feature ($x_1$ to $x_D$) and neuron ($n_1$ to $n_H$).
    }
    \label{fig:data-loader}   
    \vspace{-12.5px}
\end{figure}

Figure~\ref{fig:data-loader} shows the design of the data loader. 
Datapoints are loaded sequentially from a preloaded dataset of $R$ rows, once transition \transitionName{read_datapoint} is fired. 
For each datapoint (i.e. a row $r$ of the dataset), the values of all $D$ features are established through their respective \textit{set} blocks, where transitions \transitionName{s_nve1} and \transitionName{s_pve1} set a feature’s value to $-1$ or $+1$ respectively by adding a token to the corresponding value place. 
Alternatively, for verification, each feature can be set non-deterministically to ensure that every possible input is covered. 
These values are then passed to the inference pipeline of Section~\ref{section:methodology:inference} after transition \transitionName{features_ready}, which only becomes enabled once every feature has been set, fires and adds a token to the \transitionName{start_from_loader} place of every multiplication segment, preventing inference from starting with a partially-loaded input. 
When inference completes and adds a token to \transitionName{inference_done}, transition \transitionName{reset_features} clears the values of all $D$ features through their respective \textit{reset} blocks, where transitions \transitionName{r_nve1} and \transitionName{r_pve1} remove the token from the $-1$ or $+1$ value place respectively, before transition \transitionName{next_datapoint} advances the loader to the following datapoint.

As such, all $D$ features of a datapoint are handled concurrently, while the $R$ datapoints are processed one at a time.
In addition to marking the value place, which serves as memory for the loader to later reset the feature, each set transition produces one token in the multiplication input place of every neuron. 
Each neuron then consumes its own token and proceeds with its own inference pipeline, so that a feature, which is set only once, is distributed across all neurons.

\subsection{Inference Operations}\label{section:methodology:inference}

\begin{figure}
    \centering
    \includegraphics[width=0.825\linewidth]{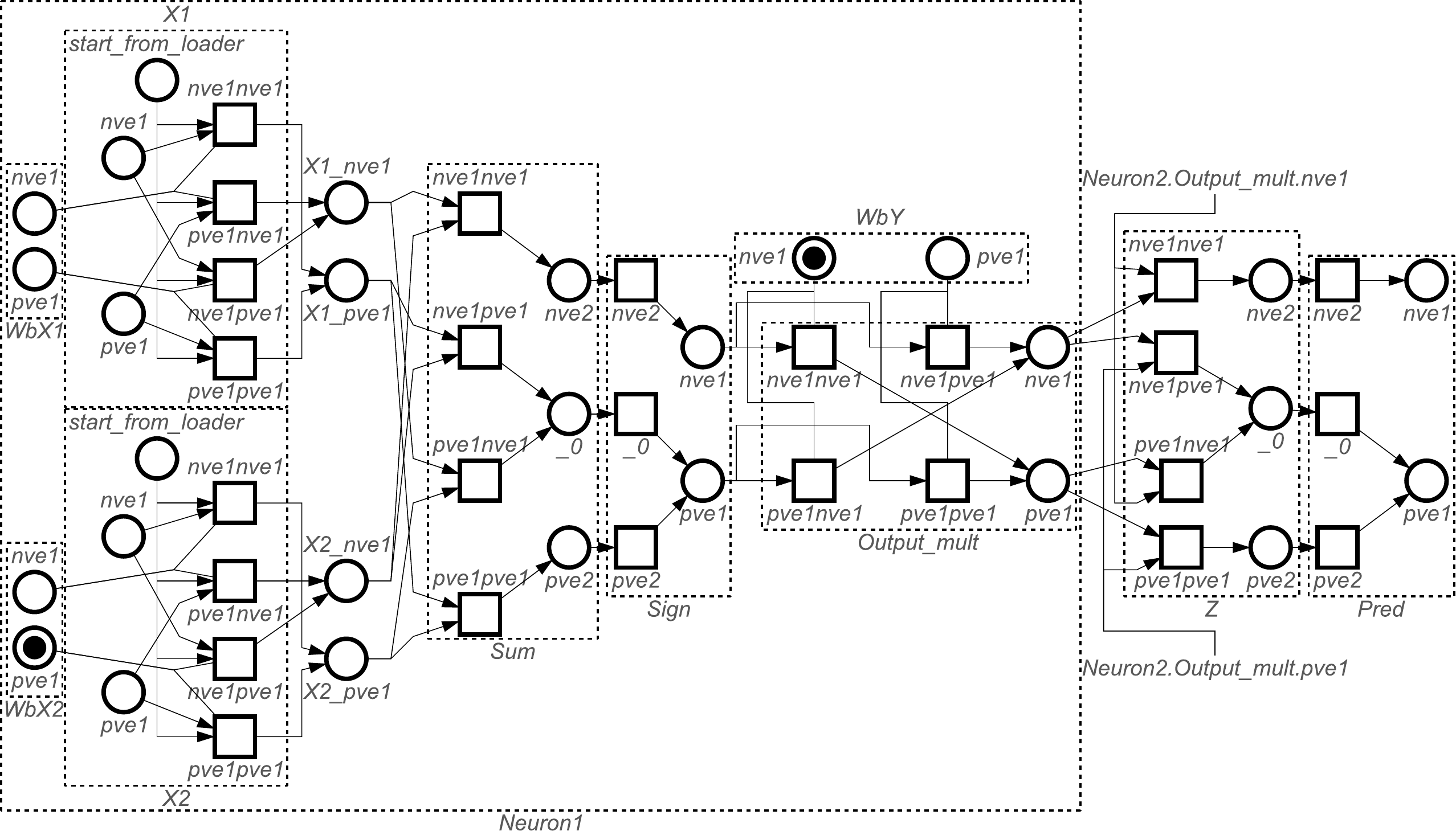}
    \caption{
        Inference pipeline for the worked example ($D = 2$) showing Neuron~1 with the output-layer aggregation ($Z$) and prediction ($\mathit{Pred}$). Neuron~2 is identical to Neuron~1 up to its weights, and labelled arc endpoints denote its output places. Transitions are named by the values they consume, e.g. \texttt{nve1pve1} in a multiplication segment fires for input $-1$ and weight $+1$, corresponding to $t_{(-1\times+1)}$ in Table~\ref{tab:trace}.
    }
    \label{fig:inference}
    \vspace{-12.5px}
\end{figure}

Inference is modelled as a pipeline of PN segments shown in Figure~\ref{fig:inference}, where the forward pass is computed via Equation~\ref{eq:forward-pass} for each input $x$.
For simplicity, this pipeline is illustrated for a two-input case, and the same structure extends to any number $D$ of features.
\vspace{-5px}
\begin{equation}
    \hat{y} = \mathrm{Sign}\!\left(\sum_{h=1}^{H} W_{bY,h} \cdot \mathrm{Sign}\!\left(\sum_{m=1}^{D} W_{bX,h,m}\, x_m\right)\right)
    \label{eq:forward-pass}
    \vspace{-5px}
\end{equation}
Note that Equation~\ref{eq:forward-pass} describes a network with a single hidden layer, where each hidden neuron sums over the $D$ input features and $H = k_2$ is the number of hidden neurons. For deeper networks, neurons in subsequent layers instead sum over the $k_{l-1}$ outputs of the previous layer.

Inference begins by multiplying each input feature by its corresponding binary weight. As multiplication cannot be easily computed arithmetically in a PN, each of the possible input–weight value combinations is represented by a dedicated transition that leads to a distinct product place. Because the BNN is trained and fixed, the weight is a constant read via a read arc meaning only transitions corresponding to that weight are enabled at inference.
Formally, this can be defined as $G_{multXm} = (P_{multXm},\ T_{multXm},\ F_{multXm},\ M_{multXm})$, where
$P_{multXm} = \{p_{x_m.k},\ p_{W_{bXm}.k},\ p_{multXm.k},\ p_{start}\}$ such that $k \in \{-1,+1\}$,
$T_{multXm} = \{t_{(-1\times-1)},\ t_{(-1\times+1)},\ t_{(+1\times-1)},\ t_{(+1\times+1)}\}$,
$F_{multXm} = \{(p_{x_m.a}, t_{(a\times b)}),\, (p_{W_{bXm}.b}, t_{(a\times b)}),$ $(p_{start}, t_{(a\times b)}) \mid a,b \in \{-1,+1\}\} \;\cup
\{(t_{(a\times b)}, p_{multXm.(a\cdot b)}),\, (t_{(a\times b)}, p_{W_{bXm}.b}) \mid a,b \in \{-1,+1\}\}$,
$M_{multXm}($ $p_{x_m.-1}\ \text{or}\ p_{x_m.+1}) = 1$,
$M_{multXm}(p_{W_{bXm}.-1}\ \text{or}\ p_{W_{bXm}.+1}) = 1$,
$M_{multXm}(p_{multXm.k}) = M_{multXm}(p_{start}) = 0$.
Here, $p_{x_m.k}$ and $p_{W_{bXm}.k}$ denote the places holding value $k$ for input $x_m$ and its weight respectively, $p_{multXm.k}$ the place holding their product, $p_{start}$ the place marked by the loader once every feature has been set, and $t_{(a\times b)}$ the transition that fires when the input and weight take values $a$ and $b$. In particular, $p_{x_m.-1}$, $p_{x_m.+1}$ and $p_{start}$ correspond to the \textit{nve1}, \textit{pve1} and \textit{start_from_loader} multiplication places respectively in Figure~\ref{fig:data-loader}.
For example, with $x_2 = +1$ and $W_{bX2} = +1$, transition $t_{(+1\times+1)}$ fires and marks the product place $p_{multX2.+1}$. 
The complete firing sequences for a worked example are given in Table~\ref{tab:trace}.

The resulting products are summed to form the pre-activation value: the summation segment $G_{sum}$ consumes the tokens from the \textit{multiplication} (\emph{mult}) output places, and produces a token at the single place encoding the pre-activation $s \in \{-2,0,+2\}$ for a two-input case, as shown in Figure~\ref{fig:inference}.
This pre-activation output is then passed through the \textit{Sign} function block to calculate the activation output, which can be formally defined as
$G_{sign} = (P_{sign},\ T_{sign},\ F_{sign},\ M_{{sign}})$ where
$P_{sign} = \{p_{sum.s},\ p_{sign.k}\}$ such that $k \in \{-1,+1\}$,
$T_{sign} = \{t_{sign(s)} \mid s \in \{-2,0,+2\}\}$,
$F_{sign} = \{(p_{sum.s},\, t_{sign(s)}),\ (t_{sign(s)},\, p_{sign.\mathrm{Sign}(s)}) \mid s \in \{-2,0,+2\}\}$,
$M_{{sign}}(p_{sum.-2}\ \text{or}\ p_{sum.0}\ \text{or}\ p_{sum.+2}) = 1,$
$M_{{sign}}(p_{sign.k}) = 0$.
Here, $t_{sign(0)}$ marks $p_{sign.+1}$, following the convention $\mathrm{Sign}(0) = +1$ defined in Section~\ref{section:background:BNN}.

Finally, the activation is multiplied by the hidden-output weight $W_{bY}$ via a multiplication segment of the same form as $G_{multXm}$, where this \textit{multiply}-\textit{sum}-\textit{Sign} pipeline is identical across all hidden neurons.

\subsection{Composed Inference Module}

The PN segments described above are composed into a full inference net by connecting each segment’s output places to the next segment’s input places, following the BNN’s feed-forward structure. 
In particular, the loader in Figure~\ref{fig:data-loader} connects to each neuron’s pipeline at the input places of its multiplication segments. 
Each hidden neuron instantiates the \textit{multiply}-\textit{sum}-\textit{Sign} pipeline shown in Figure~\ref{fig:inference}, whose output feeds the output-layer aggregation and final \textit{Sign} function that produces the prediction $\hat{y}$. 
We verify that the composed net remains 1-safe, alongside its functional correctness, in Section~\ref{section:evaluation}.

%%%%%%%%%%%%%%%%%%%%%%%%%%%%%%%%%%%%%%%%%%%%%%%%%%%%%%%%%%%%%%%%%%%%%
%%%%%%%%%%%%%%%%%%%%%%%%%%%%%%%%%%%%%%%%%%%%%%%%%%%%%%%%%%%%%%%%%%%%%
%%%%%%%%%%%%%%%%%%%%%%%%%%%%%%%%%%%%%%%%%%%%%%%%%%%%%%%%%%%%%%%%%%%%%

\vspace{-10px}
\section{Faithfulness Evaluation}\label{section:evaluation}

We validate our encoding on a trained network of two inputs, two hidden neurons, and one output, with no bias, using \mpsat{}’s \reach{} expressions in \workcraft{}. 
Here, \emph{faithfulness} means that the PN computes the same function as a reference BNN, regardless of the BNN’s accuracy. 
Both use the same binarised weights, where $Neuron1.W_{bX1} = -1$, $Neuron1.W_{bX2} = +1$, $Neuron1.W_{bY} = -1$, $Neuron2.W_{bX1} = +1$, $Neuron2.W_{bX2} = -1$, and $Neuron2.W_{bY} = -1$. 
These weights are fixed in the initial marking and read via read arcs, while, in each cycle, the loader of Section~\ref{section:methodology:input-loading} non-deterministically sets each feature to $-1$ or $+1$. 
Because 
this cycle can repeat indefinitely, 
% this PN model is \textit{reversible} (i.e. can return to its initial marking), the cycle can repeat indefinitely meaning
the reachability analysis explores every input, and every marking reachable across all cycles.

\vspace{-5px}
\subsection{Input--Output Equivalence}\label{section:validation:equivalence}

For each input $x \in \{-1,+1\}^{D}$, we executed the PN by firing the set transitions corresponding to $x$, and compared the resulting output marking against $f(x)$ from Equation~\ref{eq:forward-pass}. 
As the input space is small, we checked all $2^{D}$ inputs exhaustively, with the results shown in Table~\ref{tab:equivalence}, where the output marking encodes $f(x)$ in every case. 
We also checked for determinacy using the \reach{} expression $(M(x_1^-) \wedge M(x_2^-) \wedge M(\mathit{Pred}^+))~\vee$ $(M(x_1^+) \wedge M(x_2^-) \wedge M(\mathit{Pred}^-))~\vee (M(x_1^-) \wedge M(x_2^+) \wedge M(\mathit{Pred}^-)) \vee$ $(M(x_1^+) \wedge$ $M(x_2^+) \wedge M(\mathit{Pred}^+))$, where $x_m^{-}$ and $x_m^{+}$ denote the $-1$ and $+1$ value places of feature $x_m$ in Figure~\ref{fig:data-loader}, which persist until the features are reset, and $\mathit{Pred}^-$ and $\mathit{Pred}^+$ denote the prediction places.
The query asserting that a marking pairing any input with an incorrect output is reachable returns \textit{unsatisfiable}, which ensures no execution reaches a competing, incorrect result. 
Conversely, the corresponding query for each input’s correct output is satisfiable, which confirms that the unsatisfiable result is not vacuous. 
Together with Table~\ref{tab:equivalence}, this establishes that the PN computes the same function as the reference BNN over the input space. 
As every input is covered, this forms a complete check of faithfulness for this network rather than a sample. Establishing faithfulness for arbitrary networks, however, requires the correctness-of-translation theorem discussed in Section~\ref{section:conclusion}. 
For reference, the composed PN model for this network contains 66 places and 60 transitions and is 1-safe, with each \reach{} check completed in under 0.1 seconds.

\begin{table}[ht]
    \centering
    \footnotesize
    \vspace{-10px}
    \caption{Exhaustive input-output equivalence between the reference BNN and the PN model.}
    \vspace{5px}
    \begin{tabular}{lcccc}
       \toprule
       Input $\langle x_1,x_2\rangle$ & $\langle -1,-1\rangle$ & $\langle -1,+1\rangle$ & $\langle +1,-1\rangle$ & $\langle +1,+1\rangle$ \\
       \midrule
       Reference $f(x)$ / PN output & $-1$ / $-1$ & $+1$ / $+1$ & $+1$ / $+1$ & $-1$ / $-1$ \\
       \bottomrule
    \end{tabular}
    \vspace{-15px}
    \label{tab:equivalence}
\end{table}

\subsection{Explanation of Decision Boundary Changes through Causal Traces}\label{section:validation:trace}

Beyond establishing faithfulness, the PN encoding also allows us to explain how a decision is reached. 
Again, using the above binarised weights, we take a reference input and perturb a single feature to examine how the prediction changes. 
For the reference input $\langle -1, -1 \rangle$, the BNN classifies it as $-1$. However, flipping feature $x_2$ to give the perturbed input $\langle -1, +1 \rangle$ changes the prediction to $+1$, which forms a counterfactual explanation~\cite{wachter2017counterfactual} that the PN encoding extends by exposing the internal path through which the change takes effect.
The perturbed prediction is obtained as evidence of a reachability query asserting that the perturbed input can reach the opposite output place, which \mpsat{} returns as a firing sequence rather than only an assignment of input values.

By contrasting these two firing sequences, we localise where the decision changes, as shown in Table~\ref{tab:trace}. 
Here, the executions diverge as soon as $x_2$ is consumed, which causes different multiplication transitions to fire in both neurons and change both pre-activations.
In Neuron~1, the pre-activation rises from $0$ to $+2$ causing a different \textit{Sign} transition to fire, but yields the same activation. 
In Neuron~2, the pre-activation falls from $0$ to $-2$, so the firing \textit{Sign} transition flips its activation and, in turn, its output. 
This single neuron's flip shifts the output sum $Z$ from $-2$ to $0$, which the final \textit{Sign} resolves to the opposite prediction.
Therefore, the decision change localises to Neuron~2, whereas Neuron~1's output remains unchanged. 
In other words, at this input the network’s decision is sensitive to $x_2$ only through Neuron~2’s activation threshold.

While a constraint-based solver would only report that the perturbed input yields a different output, the firing sequence instead exposes the ordered, causal path of how the decision changed, and isolates the responsible neuron.
Similarly, observing activation values alone would only show the output of each neuron. 
By contrast, in the firing sequence, the prediction’s dependency on Neuron~2’s \textit{Sign} transition, and the absorption of the perturbation at Neuron~1’s, are explicit in which transitions fired and in what order.
This structural information is what neither constraint-based verifiers nor activation values expose, and it directly supports the fault localisation motivated in Section~\ref{section:introduction}: For an on-device classifier, an engineer can trace an unexpected decision to the responsible neuron, as a first step in evaluating whether the fault is a systematic flaw in the model or an isolated case.
We present this as a demonstration on a minimal network, with its scalability, the automation of such trace extraction, and extensions to compare distinct networks highlighted as future work in Section~\ref{section:conclusion}.

\begin{table}[ht]
    \centering
    \footnotesize
    \setlength{\tabcolsep}{4pt}
    \vspace{-20px}
    \caption{Firing sequences of the reference and perturbed inputs, grouped by causal chain. Neuron~1 and Neuron~2 chains are concurrent, while the output chain depends on both. Bold entries mark transitions that fire differently between the two runs.}
    \vspace{5px}
    \label{tab:trace}
    \begin{tabular}{llll}
        \toprule
            Chain & Operation & Reference $\langle -1,-1\rangle$ & Perturbed $\langle -1,+1\rangle$ \\
        \midrule
            \multirow{5}{*}{Neuron~1} 
              & $x_1 \times W_{bX1}$ & $t_{(-1\times-1)} \rightarrow +1$        & $t_{(-1\times-1)} \rightarrow +1$ \\
              & $x_2 \times W_{bX2}$ & $t_{(-1\times+1)} \rightarrow -1$        & $\bm{t_{(\bm{+1\times+1})} \rightarrow +1}$ \\
              & Sum                  & $t_{(+1+-1)} \rightarrow 0$              & $\bm{t_{\bm{(+1++1)}} \rightarrow +2}$ \\
              & \textit{Sign}        & $t_{\mathrm{sign}(0)} \rightarrow +1$    & $\bm{t_{\bm{\textbf{sign}(+2)}}} \rightarrow +1$ \\
              & act $\times W_{bY}$  & $t_{(+1\times-1)} \rightarrow -1$        & $t_{(+1\times-1)} \rightarrow -1$ \\
        \midrule
            \multirow{5}{*}{Neuron~2}
              & $x_1 \times W_{bX1}$ & $t_{(-1\times+1)} \rightarrow -1$        & $t_{(-1\times+1)} \rightarrow -1$ \\
              & $x_2 \times W_{bX2}$ & $t_{(-1\times-1)} \rightarrow +1$        & $\bm{t_{\bm{(+1\times-1)}} \rightarrow -1}$ \\
              & Sum                  & $t_{(-1++1)} \rightarrow 0$              & $\bm{t_{\bm{(-1+-1)}} \rightarrow -2}$ \\
              & \textit{Sign}        & $t_{\mathrm{sign}(0)} \rightarrow +1$    & $\bm{t_{\bm{\textbf{sign}(-2)}} \rightarrow -1}$ \\
              & act $\times W_{bY}$  & $t_{(+1\times-1)} \rightarrow -1$        & $\bm{t_{\bm{(-1\times-1)}} \rightarrow +1}$ \\
        \midrule
            \multirow{2}{*}{Output} 
              & $Z$ sum              & $t_{(-1+-1)} \rightarrow -2$             & $\bm{t_{\bm{(-1++1)}} \rightarrow 0}$ \\
              & \textit{Sign}        & $t_{\mathrm{sign}(-2)} \rightarrow -1$   & $\bm{t_{\bm{\textbf{sign}(0)}} \rightarrow +1}$ \\
        \bottomrule
    \end{tabular}
    \vspace{-10px}
\end{table}

%%%%%%%%%%%%%%%%%%%%%%%%%%%%%%%%%%%%%%%%%%%%%%%%%%%%%%%%%%%%%%%%%%%%%
%%%%%%%%%%%%%%%%%%%%%%%%%%%%%%%%%%%%%%%%%%%%%%%%%%%%%%%%%%%%%%%%%%%%%
%%%%%%%%%%%%%%%%%%%%%%%%%%%%%%%%%%%%%%%%%%%%%%%%%%%%%%%%%%%%%%%%%%%%%

\section{Conclusion and Future Work}\label{section:conclusion}
In this paper, we presented an early work on a discrete-event view of BNN inference by translating a fixed, trained BNN into a 1-safe PN, where each computation step is an explicit transition. On a worked example, we established faithfulness exhaustively against a reference BNN, and showed how contrasting firing sequences localises where a single-feature change alters the prediction, which exposes the causal structure that constraint-based verifiers discard.

Still, this is a first step rather than a finished method. As the encoding enumerates the values each quantity can take, it applies to discrete-valued models such as BNNs and, potentially, low-bit quantised networks. Within this scope, our prior work on Tsetlin Machines~\cite{chan2026animate} showed that automated PN construction can scale to high-dimensional datasets.
The bottleneck is instead reachability analysis, performed here by \workcraft{}’s \mpsat{} backend, which is bounded by the reachable state space and can grow considerably faster than the model itself. 
Future work therefore includes characterising this limit for BNNs, integrating BNN support into automated construction, and extending the encoding to convolutional and pooling layers and to networks whose weights evolve during training, which would require a dynamically changing PN. 
A general guarantee of faithfulness requires a correctness-of-translation theorem proved by structural induction, with the convergence of concurrent segments as a key obstacle, while the manual, single-instance causal trace would benefit from automation and an extension to localise where two trained BNNs diverge. 
Together, these would allow designers to automatically generate and query the PN model of a trained BNN, as an explanatory complement to existing BNN verification for the dependable deployment of BNNs within larger systems.

% %%%%%%%%%%%%%%%%%%%%%%%%%%%%%%%%%%%%%%%%%%%%%%%%%%%%%%%%%%%%%%%%%%%%%
% %%%%%%%%%%%%%%%%%%%%%%%%%%%%%%%%%%%%%%%%%%%%%%%%%%%%%%%%%%%%%%%%%%%%%
% %%%%%%%%%%%%%%%%%%%%%%%%%%%%%%%%%%%%%%%%%%%%%%%%%%%%%%%%%%%%%%%%%%%%%

\vspace{-12.5px}
\section*{Acknowledgments}

This work was supported by EPSRC EP/X036006/1 Scalability Oriented Novel Network of Event Triggered Systems (SONNETS) project.

%%%%%%%%%%%%%%%%%%%%%%%%%%%%%%%%%%%%%%%%%%%%%%%%%%%%%%%%%%%%%%%%%%%%%
%%%%%%%%%%%%%%%%%%%%%%%%%%%%%%%%%%%%%%%%%%%%%%%%%%%%%%%%%%%%%%%%%%%%%
%%%%%%%%%%%%%%%%%%%%%%%%%%%%%%%%%%%%%%%%%%%%%%%%%%%%%%%%%%%%%%%%%%%%%

\vspace{-12.5px}
\section*{AI Usage}

Following the guidelines of FMAS 2026, the authors disclose the use of the large language model, Anthropic Claude, for editing and improving the grammar of this paper. 

%%%%%%%%%%%%%%%%%%%%%%%%%%%%%%%%%%%%%%%%%%%%%%%%%%%%%%%%%%%%%%%%%%%%%
%%%%%%%%%%%%%%%%%%%%%%%%%%%%%%%%%%%%%%%%%%%%%%%%%%%%%%%%%%%%%%%%%%%%%
%%%%%%%%%%%%%%%%%%%%%%%%%%%%%%%%%%%%%%%%%%%%%%%%%%%%%%%%%%%%%%%%%%%%%

\bibliographystyle{eptcs}
\bibliography{references}

\end{document}